\pdfoutput=1

\documentclass[12pt,a4paper]{article}

\usepackage{ifthen} 
\newboolean{pdflatex}
\setboolean{pdflatex}{true} 

\newboolean{articletitles}
\setboolean{articletitles}{true} 

\newboolean{uprightparticles}
\setboolean{uprightparticles}{false} 

\newboolean{inbibliography}
\setboolean{inbibliography}{false} 

\usepackage{microtype}
\usepackage{lineno}  
\usepackage{xspace} 

\usepackage{graphicx}  
\usepackage{color}
\usepackage{colortbl}
\graphicspath{{./figs/}} 

\usepackage{amsmath} 
\usepackage{amssymb}
\usepackage{amsfonts}
\usepackage{upgreek} 

\newcommand*\patchAmsMathEnvironmentForLineno[1]{%
\expandafter\let\csname old#1\expandafter\endcsname\csname #1\endcsname
\expandafter\let\csname oldend#1\expandafter\endcsname\csname
end#1\endcsname
 \renewenvironment{#1}%
   {\linenomath\csname old#1\endcsname}%
   {\csname oldend#1\endcsname\endlinenomath}%
}
\newcommand*\patchBothAmsMathEnvironmentsForLineno[1]{%
  \patchAmsMathEnvironmentForLineno{#1}%
  \patchAmsMathEnvironmentForLineno{#1*}%
}
\AtBeginDocument{%
\patchBothAmsMathEnvironmentsForLineno{equation}%
\patchBothAmsMathEnvironmentsForLineno{align}%
\patchBothAmsMathEnvironmentsForLineno{flalign}%
\patchBothAmsMathEnvironmentsForLineno{alignat}%
\patchBothAmsMathEnvironmentsForLineno{gather}%
\patchBothAmsMathEnvironmentsForLineno{multline}%
}

\usepackage{hyperref}    
\usepackage[all]{hypcap} 

\def\lhcb {\mbox{LHCb}\xspace}
\def\ux85 {\mbox{UX85}\xspace}

\ifthenelse{\boolean{uprightparticles}}%
{

 \def\Ppsi        {\ensuremath{\uppsi}\xspace}

 \def\PDelta      {\ensuremath{\Delta}\xspace}                 
 \def\PXi      {\ensuremath{\Xi}\xspace}                 
 \def\PLambda      {\ensuremath{\Lambda}\xspace}                 
 \def\PSigma      {\ensuremath{\Sigma}\xspace}                 
 \def\POmega      {\ensuremath{\Omega}\xspace}                 
 \def\PUpsilon      {\ensuremath{\Upsilon}\xspace}                 
 
 \def\PB      {\ensuremath{\mathrm{B}}\xspace}                 
                  
 \def\PD      {\ensuremath{\mathrm{D}}\xspace}

 \def\PJ      {\ensuremath{\mathrm{J}}\xspace}                 
 \def\PK      {\ensuremath{\mathrm{K}}\xspace}

 \def\Pb      {\ensuremath{\mathrm{b}}\xspace}                 
 \def\Pc      {\ensuremath{\mathrm{c}}\xspace}

 \def\Pi      {\ensuremath{\mathrm{i}}\xspace}

 \def\Ps      {\ensuremath{\mathrm{s}}\xspace}

}
{

 \def\Ppsi        {\ensuremath{\psi}\xspace}                 
                  
 \mathchardef\PDelta="7101
 \mathchardef\PXi="7104
 \mathchardef\PLambda="7103
 \mathchardef\PSigma="7106
 \mathchardef\POmega="710A
 \mathchardef\PUpsilon="7107
                  
 \def\PB      {\ensuremath{B}\xspace}                 
                  
 \def\PD      {\ensuremath{D}\xspace}

 \def\PJ      {\ensuremath{J}\xspace}                 
 \def\PK      {\ensuremath{K}\xspace}

 \def\Pb      {\ensuremath{b}\xspace}                 
 \def\Pc      {\ensuremath{c}\xspace}

 \def\Pi      {\ensuremath{i}\xspace}

 \def\Ps      {\ensuremath{s}\xspace}

}

\def\squark    {\ensuremath{\Ps}\xspace}

\def\cquark    {\ensuremath{\Pc}\xspace}

\def\bquark    {\ensuremath{\Pb}\xspace}

\def\kaon  {\ensuremath{\PK}\xspace}
  \def\Kbar  {\kern 0.2em\overline{\kern -0.2em \PK}{}\xspace}

\def\Kz    {\ensuremath{\kaon^0}\xspace}
\def\Kzb   {\ensuremath{\Kbar^0}\xspace}
\def\KzKzb {\ensuremath{\Kz \kern -0.16em \Kzb}\xspace}
\def\Kp    {\ensuremath{\kaon^+}\xspace}
\def\Km    {\ensuremath{\kaon^-}\xspace}

\def\KpKm  {\ensuremath{\Kp \kern -0.16em \Km}\xspace}

\def\Kstarzb {\ensuremath{\Kbar^{*0}}\xspace}

  \def\Dbar    {\kern 0.2em\overline{\kern -0.2em \PD}{}\xspace}
\def\D       {\ensuremath{\PD}\xspace}

\def\Dz      {\ensuremath{\D^0}\xspace}
\def\Dzb     {\ensuremath{\Dbar^0}\xspace}
\def\DzDzb   {\ensuremath{\Dz {\kern -0.16em \Dzb}}\xspace}
\def\Dp      {\ensuremath{\D^+}\xspace}
\def\Dm      {\ensuremath{\D^-}\xspace}

\def\DpDm    {\ensuremath{\Dp {\kern -0.16em \Dm}}\xspace}

\def\B       {\ensuremath{\PB}\xspace}
\def\Bbar    {\ensuremath{\kern 0.18em\overline{\kern -0.18em \PB}{}}\xspace}

\def\Bzb     {\ensuremath{\Bbar^0}\xspace}

\def\Bs      {\ensuremath{\B^0_\squark}\xspace}
\def\Bsb     {\ensuremath{\Bbar^0_\squark}\xspace}
\def\Bdb     {\ensuremath{\Bbar^0}\xspace}

\def\jpsi     {\ensuremath{{\PJ\mskip -3mu/\mskip -2mu\Ppsi\mskip 2mu}}\xspace}

  \def\Y#1S{\ensuremath{\PUpsilon{(#1S)}}\xspace}

\def\L {\ensuremath{\PLambda}\xspace}
\def\Lz {\ensuremath{\PLambda}\xspace}
\def\Lbar {\ensuremath{\kern 0.1em\overline{\kern -0.1em\PLambda}}\xspace}

\def\Lb      {\ensuremath{\L^0_\bquark}\xspace}

\def\Lc      {\ensuremath{\L^+_\cquark}\xspace}

\def\to                 {\ensuremath{\rightarrow}\xspace}

\def\CP                {\ensuremath{C\!P}\xspace}

\def\AT#1     {\ensuremath{A_{\mathrm{T}}^{#1}}\xspace}           

\def\C#1      {\ensuremath{\mathcal{C}_{#1}}\xspace}                       
\def\Cp#1     {\ensuremath{\mathcal{C}_{#1}^{'}}\xspace}                    
\def\Ceff#1   {\ensuremath{\mathcal{C}_{#1}^{\mathrm{(eff)}}}\xspace}        
\def\Cpeff#1  {\ensuremath{\mathcal{C}_{#1}^{'\mathrm{(eff)}}}\xspace}       
\def\Ope#1    {\ensuremath{\mathcal{O}_{#1}}\xspace}                       
\def\Opep#1   {\ensuremath{\mathcal{O}_{#1}^{'}}\xspace}                    

\newcommand{\tev}{\ifthenelse{\boolean{inbibliography}}{\ensuremath{~T\kern -0.05em eV}\xspace}{\ensuremath{\mathrm{\,Te\kern -0.1em V}}\xspace}}
\newcommand{\gev}{\ensuremath{\mathrm{\,Ge\kern -0.1em V}}\xspace}
\newcommand{\mev}{\ensuremath{\mathrm{\,Me\kern -0.1em V}}\xspace}
\newcommand{\kev}{\ensuremath{\mathrm{\,ke\kern -0.1em V}}\xspace}
\newcommand{\ev}{\ensuremath{\mathrm{\,e\kern -0.1em V}}\xspace}
\newcommand{\gevc}{\ensuremath{{\mathrm{\,Ge\kern -0.1em V\!/}c}}\xspace}
\newcommand{\mevc}{\ensuremath{{\mathrm{\,Me\kern -0.1em V\!/}c}}\xspace}
\newcommand{\gevcc}{\ensuremath{{\mathrm{\,Ge\kern -0.1em V\!/}c^2}}\xspace}
\newcommand{\gevgevcccc}{\ensuremath{{\mathrm{\,Ge\kern -0.1em V^2\!/}c^4}}\xspace}
\newcommand{\mevcc}{\ensuremath{{\mathrm{\,Me\kern -0.1em V\!/}c^2}}\xspace}

\def\gsim{{~\raise.15em\hbox{$>$}\kern-.85em
          \lower.35em\hbox{$\sim$}~}\xspace}
\def\lsim{{~\raise.15em\hbox{$<$}\kern-.85em
          \lower.35em\hbox{$\sim$}~}\xspace}

\def\pt         {\mbox{$p_{\rm T}$}\xspace}

\def\gauss      {\mbox{\textsc{Gauss}}\xspace}

\def\tell1  {TELL1\xspace}
\def\ukl1   {UKL1\xspace}

\usepackage{cite} 
\usepackage{mciteplus}

\usepackage{longtable} 

\begin{document}

\renewcommand{\thefootnote}{\fnsymbol{footnote}}
\setcounter{footnote}{1}

\begin{titlepage}
\pagenumbering{roman}

\vspace*{-1.5cm}
\centerline{\large EUROPEAN ORGANIZATION FOR NUCLEAR RESEARCH (CERN)}
\vspace*{1.5cm}
\hspace*{-5mm}\begin{tabular*}{16cm}{lc@{\extracolsep{\fill}}r}
\vspace*{-12mm}\mbox{\!\!\!\includegraphics[width=.12\textwidth]{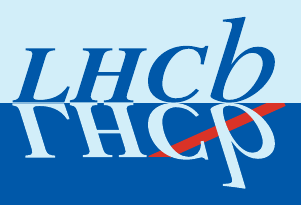}}& & \\ 
 & & CERN-PH-EP-2013-117\\
 & & LHCb-PAPER-2013-032\\  
 & & July 9, 2013 \\ 
 & & \\
\end{tabular*}

\vspace*{3.0cm}

{\bf\boldmath\huge
\begin{center}
  Precision measurement of the \Lb baryon lifetime
\end{center}
}

\vspace*{2.0cm}

\begin{center}
The LHCb collaboration\footnote{Authors are listed on the following pages.}
\end{center}

\vspace{\fill}

\begin{abstract}
  \noindent
The ratio of the  \Lb baryon lifetime to that 
of the \Bzb meson  is measured using 1.0~fb$^{-1}$ of integrated luminosity in 7 TeV center-of-mass energy $pp$ collisions at the LHC.  The \Lb baryon is observed for the first time in the decay mode  $\Lb\to\jpsi p K^-$, while the \Bzb meson decay used is the well known 
 $\Bzb\to\jpsi \pi^+ K^-$ mode, where the $\pi^+K^-$ mass is consistent with that of the $\Kstarzb(892)$ meson. The  ratio of lifetimes  is measured to be $0.976\pm0.012\pm0.006$, in agreement with theoretical expectations based on the heavy quark expansion. Using previous determinations of the \Bzb meson lifetime, the \Lb lifetime is found to be $1.482 \pm 0.018 \pm 0.012$~ps. In both cases the first uncertainty is statistical and the second systematic.
\end{abstract}

\vspace*{2.0cm}

\begin{center}
  Submitted to Phys.~Rev.~Lett. 
\end{center}

\vspace{\fill}

{\footnotesize 
\centerline{\copyright~CERN on behalf of the \lhcb collaboration, license \href{http://creativecommons.org/licenses/by/3.0/}{CC-BY-3.0}.}}
\vspace*{2mm}

\end{titlepage}

\newpage
\setcounter{page}{2}
\mbox{~}

\centerline{\large\bf LHCb collaboration}
\begin{flushleft}
\small
R.~Aaij$^{40}$, 
B.~Adeva$^{36}$, 
M.~Adinolfi$^{45}$, 
C.~Adrover$^{6}$, 
A.~Affolder$^{51}$, 
Z.~Ajaltouni$^{5}$, 
J.~Albrecht$^{9}$, 
F.~Alessio$^{37}$, 
M.~Alexander$^{50}$, 
S.~Ali$^{40}$
% ...

{\footnotesize \it
$ ^{1}$Centro Brasileiro de Pesquisas F\'{i}sicas (CBPF), Rio de Janeiro, Brazil\\
% ...
$ ^{5}$Clermont Universit\'{e}, Universit\'{e} Blaise Pascal, CNRS/IN2P3, LPC, Clermont-Ferrand, France\\
$ ^{6}$CPPM, Aix-Marseille Universit\'{e}, CNRS/IN2P3, Marseille, France\\
% ...
$ ^{9}$Fakult\"{a}t Physik, Technische Universit\"{a}t Dortmund, Dortmund, Germany\\
% ...
$ ^{36}$Universidad de Santiago de Compostela, Santiago de Compostela, Spain\\
$ ^{37}$European Organization for Nuclear Research (CERN), Geneva, Switzerland\\
% ...
$ ^{40}$Nikhef National Institute for Subatomic Physics, Amsterdam, The Netherlands\\
% ...
$ ^{45}$H.H. Wills Physics Laboratory, University of Bristol, Bristol, United Kingdom\\
% ...
$ ^{50}$School of Physics and Astronomy, University of Glasgow, Glasgow, United Kingdom\\
$ ^{51}$Oliver Lodge Laboratory, University of Liverpool, Liverpool, United Kingdom\\
% ...
}
\end{flushleft}

\cleardoublepage

\renewcommand{\thefootnote}{\arabic{footnote}}
\setcounter{footnote}{0}



\pagestyle{plain} 
\setcounter{page}{1}
\pagenumbering{arabic}  

\noindent Evaluations from experimental data of fundamental parameters, such as CKM matrix elements \cite{Kowa-Mannel,*Bernlochner:2010zz}, and limits on physics beyond that described by the standard model, often rely on theoretical input  \cite{Laiho:2012ss,*Stone:2012yr}. One of the most useful models, the heavy quark expansion (HQE) \cite{Bigi:1995jr,*Bigi:1994wa,Uraltsev:1998bk,Neubert:1997gu}, is based on
the operator product expansion \cite{Wilson:1972ee,*Buchalla:1995vs}; it
is used, for example, to extract values for $|V_{ub}|$ and $|V_{cb}|$ from measurements of inclusive semileptonic $B$ meson decays \cite{Falk:2000tx,*Buras:2011we}.  
In the free quark model the lifetimes of all $b$-flavored hadrons are equal, because the decay width is determined by the $b$ quark lifetime. This model is too na\"ive, since effects of other quarks in the hadron are not taken into account \cite{Cheng:1997xba}.
Early predictions using the HQE, however, supported the idea that $b$-hadron lifetimes were quite similar, due to the absence
of correction terms ${\cal{O}}(1/m_b)$. In the case of the ratio of lifetimes of the \Lb baryon, $\tau_{\Lb}$, to the \Bzb meson, $\tau_{\Bzb}$,
the corrections of order ${\cal{O}}(1/m^2_b)$ were found to be small,  initial estimates of ${\cal{O}}(1/m^3_b)$ \cite{Neubert:1996we,Uraltsev:1996ta,*DiPierro:1999tb}  effects were also small, thus differences of only a few percent were expected \cite{Neubert:1996we,Cheng:1997xba,Rosner:1996fy}.
Measurements at LEP, however,  indicated that $\tau_{\Lb}/\tau_{\Bzb}$ was lower: in 2003 one widely quoted average of all data
gave $0.798\pm0.052$ \cite{Battaglia:2003in}, while another gave  $0.786\pm0.034$ \cite{Tarantino:2003qw,*Franco:2002fc}. 
Some authors sought to explain the small value of the ratio by including additional operators or other modifications \cite{Ito:1997qq,*Gabbiani:2003pq,*Gabbiani:2004tp}, while some thought that the
HQE could be pushed to provide a ratio of $\sim$0.9 \cite{Uraltsev:2000qw}.  Recent measurements have shown indications that a higher value is possible \cite{Aad:2012bpa,*Chatrchyan:2013sxa,*Aaltonen:2009zn,*Aaltonen:2010pj,*Abazov:2012iy}, although the uncertainties are still large. Therefore, a
precision measurement of $\tau_{\Lb}/\tau_{\Bzb}$ is necessary to provide a confirmation of the HQE, or show definitively that the theory is deficient.

In this Letter we present the experimental determination of $\tau_{\Lb}/\tau_{\Bzb}$  using a data sample corresponding to 1.0~fb$^{-1}$ of integrated luminosity accumulated by the LHCb experiment in 7~TeV center-of-mass energy $pp$ collisions. The \Lb baryon is detected in the $\jpsi p K^-$ decay mode, while the \Bzb meson is found in $\jpsi \pi^+K^-$ decays. Mention of a particular decay channel implies the additional use of the charge-conjugate mode. This \Lb decay mode has not been observed before.\footnote{Measurement of the branching fraction is under study, and will be reported in a subsequent publication.} On the other hand, the \Bzb decay is well known,  and we impose the further requirement that the invariant mass of the $\pi^+K^-$ combination be within $\pm$100~MeV of the $\Kstarzb(892)$ mass,\footnote{We work in units where $c=1$.} in order to simplify the simulation and reduce systematic uncertainties. These decays have  the same decay topology into four charged tracks, thus facilitating the cancellation of uncertainties.

The LHCb detector  is a single-arm forward
spectrometer covering the pseudorapidity range $2<\eta <5$, described in detail in Ref.~\cite{LHCb-det}.
Events selected for this analysis are triggered \cite{Aaij:2012me}  by a $J/\psi\to\mu^+\mu^-$ decay, where the   
$J/\psi$ is required at the software level to be consistent with coming from the decay of a $b$-hadron by use either of IP requirements or detachment of the $J/\psi$ from the associated primary vertex. The simulated events used in this analysis are produced using the software described in Refs.~\cite{Sjostrand:2006za,*LHCb-PROC-2011-005,*Agostinelli:2002hh,*Allison:2006ve,*LHCb-PROC-2011-006,*Lange:2001uf}


 
Events are preselected and then are further filtered using a multivariate analyzer based on the boosted decision tree (BDT) technique~\cite{Breiman}.  In the preselection,  all hadron track candidates are required to have \pt larger than 250~MeV, 
while for muon candidates the requirement is more than 550~MeV. Events must have a $\mu^+\mu^-$ combination that forms a common vertex with $\chi^2 < 16$, and an invariant mass between $-48$ and +43 MeV of the \jpsi mass. Candidate $\mu^+\mu^-$ combinations are then constrained to the \jpsi mass for subsequent use in event selection.  The two charged final state hadrons must have a vector summed \pt of more than 1 GeV, and are also required to form a vertex with $\chi^2<10$ for one degree of freedom, and a common vertex with the \jpsi candidate with $\chi^2<50$ for five degrees of freedom.  
This  $b$-hadron candidate must have a momentum vector that, when parity inverted, points to the primary vertex within an angle smaller than 2.56$^\circ$.  Particle identification requirements differ in the two modes. We use the difference in the logarithm of the likelihood, DLL$(h_1-h_2)$, to distinguish between the two hypotheses: $h_1$ and $h_2$ as described in \cite{Adinolfi:2012an}. In the \Lb decay the kaon candidate
must have DLL$(K-\pi)>4$ and DLL$(K-p)>-3$, while the proton must have DLL$(p-\pi)>10$ and DLL$(p-K)>-3$.
For the \Bzb decay, the requirements on the pion candidate are DLL$(\pi-\mu)>-10$ and DLL$(\pi-K)>-10$, while DLL$(K-\pi)>0$ is required for the kaon. 

The BDT selection is based on the minimum DLL($\mu-\pi$) of the $\mu^+$ and $\mu^-$ candidates, the $\pt$ of each of the two charged hadrons,  and their sum,  the \Lb \pt, the \Lb vertex $\chi^2$,  and the impact parameter $\chi^2$ of the \Lb candidate, where the latter results from calculating the difference in $\chi^2$  by using the hypothesis that the IP is zero.
These variables are chosen with the aim of having the selection efficiency be independent of decay time. The BDT is trained on a simulated sample of  either $\Lb\to\jpsi p K^-$ signal events and a background data sample from the mass sidebands of the \Lb  signal peak. It is then tested on independent samples from the same sources. The BDT selection is implemented to maximize $S^2/(S+B)$, where $S$ indicates the signal and $B$ the background event yields. This optimization includes the requirement that the \Lb baryon decay time be greater than 0.5~ps. The same BDT selection is used for the $\Bzb\to\jpsi \pi^-K^+$ mode.

\begin{figure}[htb]
\begin{center}
    \includegraphics[width=0.78\textwidth]{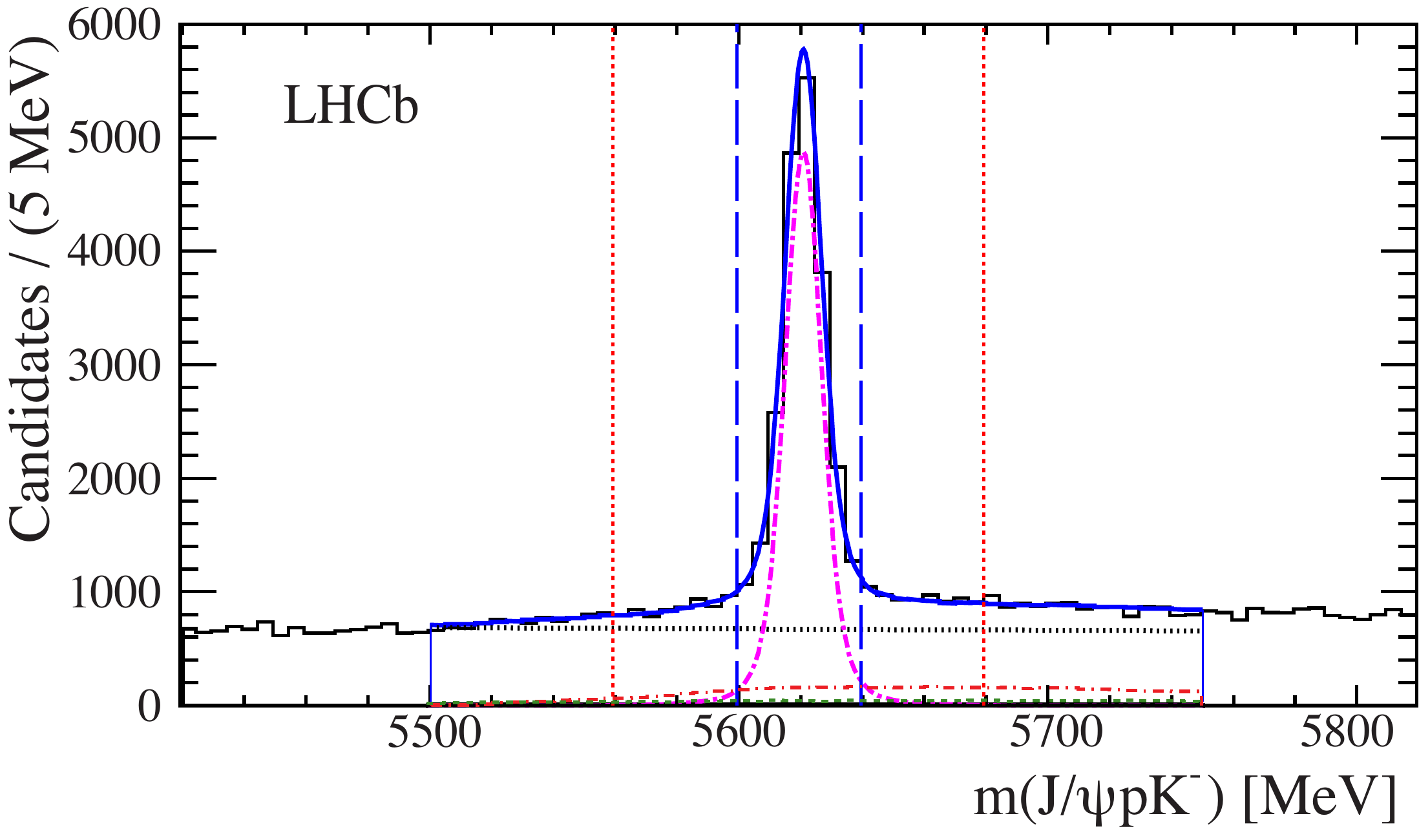}%
\end{center}\label{fig:psi-pK-mass}
\vskip -0.7cm
\caption{\small  Invariant mass spectrum of $\jpsi pK^-$ combinations. The signal region is between the vertical long dashed (blue) lines.  The sideband regions extend from the dotted (red) lines to the edges of the plot. The fit to the data between 5500 and 5750 MeV  is  also shown by the (blue) solid curve, with the $\Lb$ signal shown by the dashed-dot (magenta) curve. The  (black) dotted line is the combinatorial background and  $\Bsb\to \jpsi \KpKm$ and $\Bdb \to \jpsi \pi^+K^-$ reflections are shown with the (red) dashed-dot-dot  and (green) dashed shapes, respectively.}
\end{figure}

The $J/\psi p K^-$ mass distribution after the BDT selection is shown in Fig.~\ref{fig:psi-pK-mass}.
There is a large  and significant signal. Backgrounds can be combinatorial in nature, but can also be formed 
by reflections from $B$ meson decays where the particle identification fails.  As long as these backgrounds do not peak near the \Lb mass they cannot cause incorrect determinations of the \Lb signal yield. The shapes of the main $B$ meson reflections are determined from simulation and shown on Fig.~\ref{fig:psi-pK-mass}. The shapes are smooth and do not peak in the signal region.  
 To estimate the contributions of the reflections we take each of the candidates in the $\jpsi pK^-$ sideband regions $60-200$ MeV on either side of the \Lb mass peak, reassign proton to kaon and pion mass hypotheses, respectively, and fit the resulting signal peaks determining signal yields of  $5576\pm 95$ \Bsb and  $1769\pm192$ \Bzb decays. 
To translate these yields to those within $\pm$20 MeV of the \Lb peak, we use simulations of $\Bsb\to\jpsi K^+K^-$ with the $K^+K^-$ mass distribution matched to that obtained in our previous analysis of this final state \cite{Aaij:2013orb}, and a simulation of $\Bzb\to\jpsi \pi^+ K^-$ decays, leading to 1186$\pm$35 $\jpsi K^+K^-$ and 308$\pm$33 $\jpsi \pi^+ K^-$ reflected decays, respectively.

To determine the $\Lb$ signal yield we perform an unbinned maximum likelihood fit to the $\jpsi pK^-$ invariant mass spectrum shown in Fig.~\ref{fig:psi-pK-mass} in the region between 5500 and 5750 MeV. 
The fit function is the sum of the $\Lb$ signal component, combinatorial background and the contribution from the $\Bsb\to\jpsi \KpKm$ and $\Bdb \to \jpsi \pi^+K^-$ reflections. The signal is modeled by a triple-Gaussian function with common means; the effective r.m.s. width is 5.5~MeV. The combinatorial background is described by an exponential function. The event yields of the reflections are included in the fit as Gaussian constraints. The mass fit gives $15\,581\pm 178$ signal and $5535\pm 50$ combinatorial background candidates together with $1235\pm 35$  $\Bsb\to\jpsi \KpKm$ and $313\pm 26$ $\Bdb \to \jpsi \pi^+K^-$ reflection candidates  within $\pm 20$ MeV of the $\Lb$ mass peak.

\begin{figure}[b]
\begin{center}
    \includegraphics[width=0.70\textwidth]{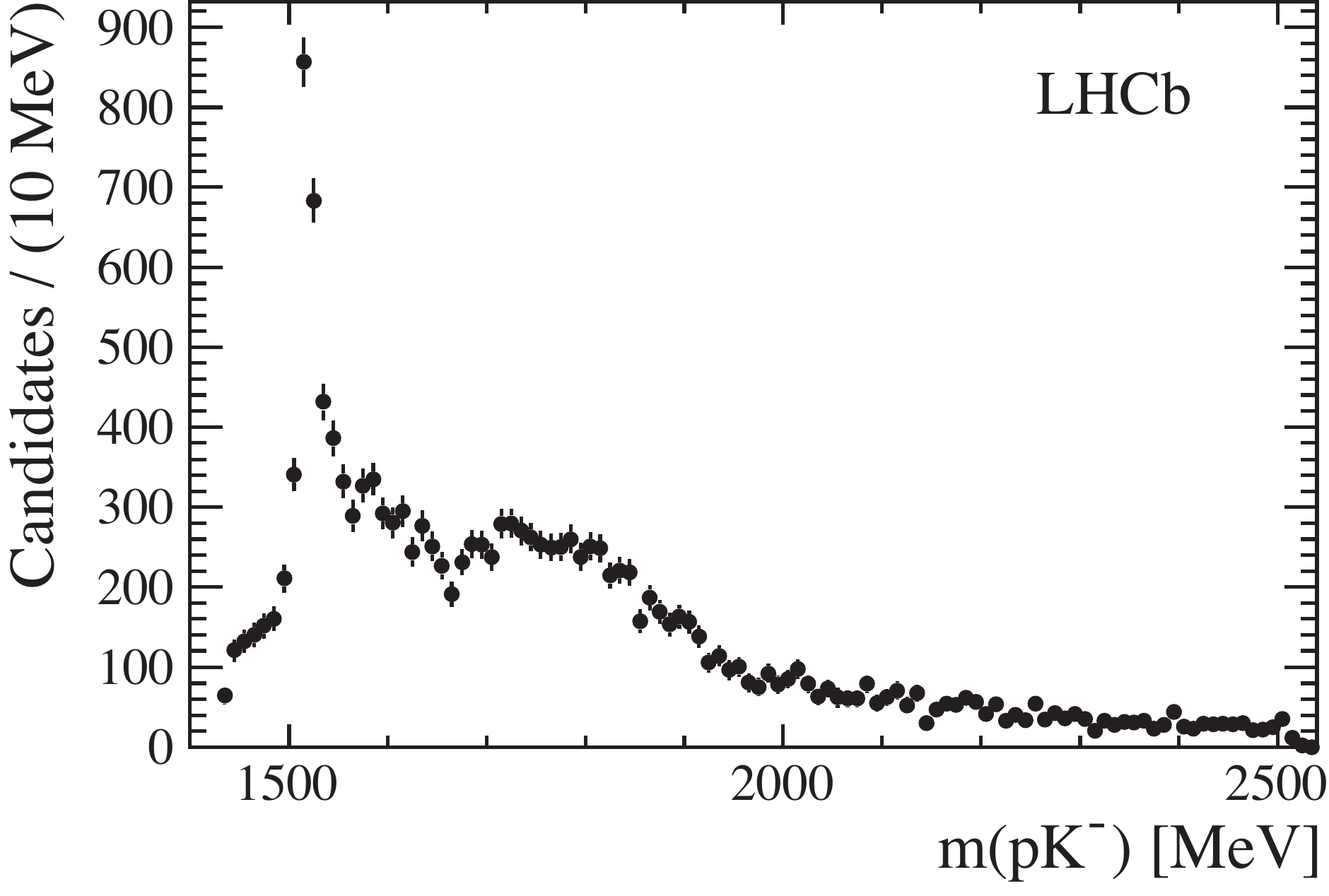}%
\end{center}\label{fig:mpK}
\vskip -0.5cm
\caption{\small Background subtracted $m(pK^-)$ distribution obtained by fitting the $m(\jpsi pK^-)$ distribution in bins of $m(pK^-)$.}
\end{figure}

To view the background subtracted $pK^-$ mass spectrum, we perform fits, as described above, to the $m(\jpsi pK^-)$ distributions in bins of $m(pK^-)$ and extract the signal yields within $\pm 20$ MeV of  the $\Lb$ mass peak. The resulting  $pK^-$ mass spectrum is shown in Fig.~\ref{fig:mpK}. 
A distinct peak is observed in the $pK^-$ invariant mass distribution near 1520 MeV, together with the other resonant and non-resonant structures over the entire kinematical region. The peak corresponds to the $\Lz(1520)$ resonance \cite{PDG}.  Simulations of the \Lb decay are weighted to reproduce this mass distribution.

\begin{figure}[h!bt]
\begin{center}
\includegraphics[width=0.70\textwidth]{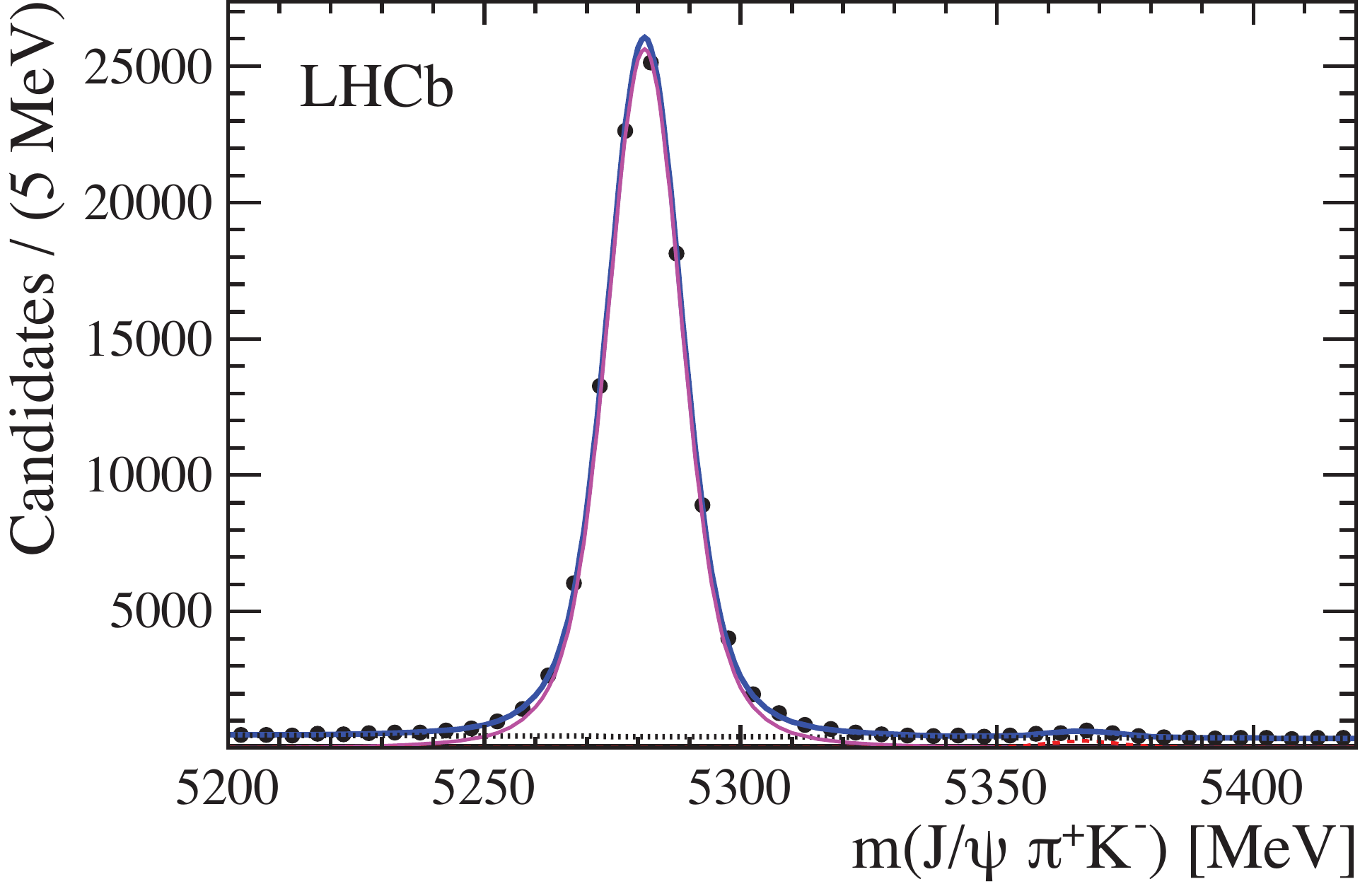}
\end{center}\label{fig:Bd2JpsiKst}
\vskip -0.4cm
\caption{\small Fit to the invariant mass spectrum of $\jpsi \pi^+K^-$ combinations with $\pi^+K^-$ invariant mass within $\pm$100~MeV of the $\Kstarzb$ mass. The $\Bdb$ signal is shown by the (magenta) solid  curve,  the combinatorial background by the (black) dotted line, the  $\Bs\to \jpsi \pi^+K^-$ signal  by the (red) dashed  curve,  and the total by the (blue) solid  curve.}
\end{figure}

The $\jpsi \pi^+K^-$ mass spectrum, after the BDT selection, is shown in Fig.~\ref{fig:Bd2JpsiKst}.  There is a large signal peak at the \Bzb mass and a much smaller one at the \Bs mass. Triple-Gaussian functions each with common means are used to fit the signal peaks; the effective r.m.s. width is 6.7~MeV.  An exponential function is used to fit the combinatorial background. The mass fit gives $97\,506\pm 447$ signal and $3660 \pm 74$ background candidates within $\pm 20 $ MeV of the $\Bdb$ mass peak. Reflections are possible from both $\Bsb\to\jpsi K^+K^-$ and $\Lb\to\jpsi p K^-$ decays. Following the same procedure as outlined above using the sidebands of the $\Bzb$ signal we find no evidence of a reflection from the $\Bsb$ state and a small, non-peaking, contribution of 506$\pm$19 events from the \Lb state, in the \Bzb signal region,  that is ignored. 

The  decay time for each candidate is given by $t =m \vec{d} \cdot \vec{p}/|\vec{p}|^2$,  where $m$ is the mass, $\vec{d}$ the distance vector from the primary vertex to the decay point, and $\vec{p}$ is the measured $b$ hadron momentum. Here, we do not constrain the two muons to the \jpsi mass to avoid systematic biases.  The decay time resolutions are 40~fs for the \Lb decay and 37~fs for the \Bzb decay.  In addition, the decay time acceptances are also almost equal. For equal acceptances, the ratio of events, $R(t)$, as a function of decay time is given by
\begin{equation} 
\label{eq:raterat}
R(t)=\frac{N_{\Lb}(0)}{N_{\Bzb}(0)}\frac{e^{-t/\tau_{\Lb}}}{e^{-t/\tau_{\Bzb}}}=R(0)e^{-t\Delta_{\Lz B}},
\end{equation}
 where $\Delta_{\Lz§ B}=\left(1/\tau_{\Lb}-1/\tau_{\Bzb}\right)$. Effects of the different decay time resolutions in the two modes are negligible above 0.5 ps. 
 First order corrections for a decay time dependent acceptance ratio can be taken into account by modifying Eq.~(\ref{eq:raterat}) with a linear function
 \begin{equation} 
\label{eq:raterat2}
R(t)=R(0)[1+a\cdot t]e^{-t\Delta_{\Lz B}},
\end{equation}
where $a$ represents the slope of the acceptance ratio as a function of  decay time.

The decay time acceptances for both modes are determined by simulations that are weighted to match either the $pK^-$ or $\pi^+K^-$ invariant mass distributions seen in data, as well as to match the measured $p$ and \pt distributions of the $b$ hadrons. In addition, 
we further weight the samples so that the simulation matches the hadron identification efficiencies  obtained from $D^{*+}\to \pi^+(D^0\to \pi^+K^-)$ events for pions and kaons, and $\L^0 \to p\pi^-$ for protons.

\begin{figure}[hbt]
\begin{center}
    \includegraphics[width=0.68\textwidth]{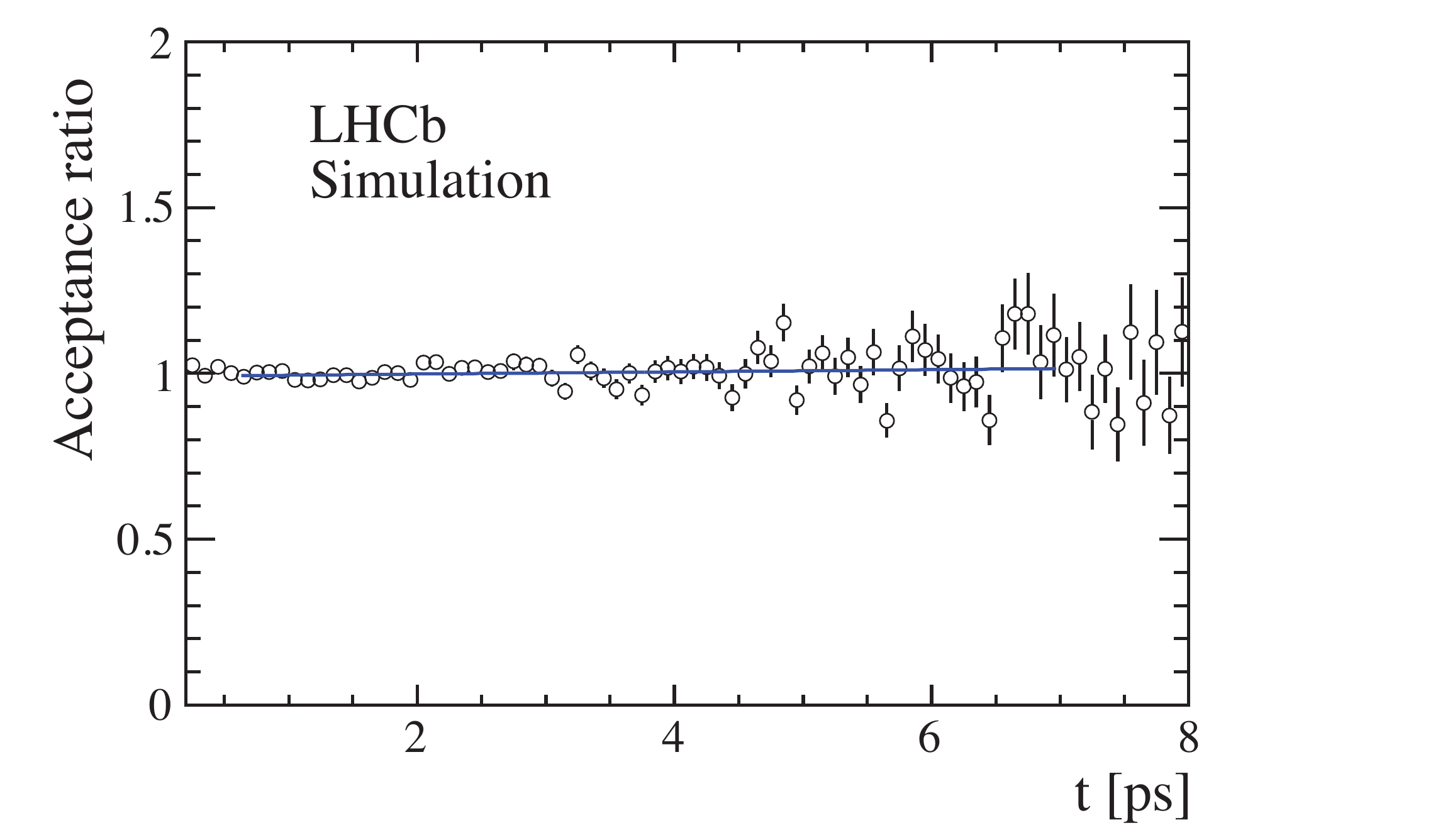}%
\end{center}\label{fig:acceptance_ratio}
\vskip -0.5cm
\caption{\small Ratio of  the decay time acceptances between $\Lb \to \jpsi pK^-$ and  $\Bdb \to \jpsi \Kstarzb(892)$ decays obtained from simulation. The (blue) line shows the result of the linear fit. }
\end{figure}

The ratio of the decay time acceptances is shown in Fig.~\ref{fig:acceptance_ratio}. Here we have removed the minimum requirement on  decay time so we can view the distributions in the region close to zero time.
The individual acceptances in both cases can be described with a  linear function above 0.5~ps. In order to minimize possible systematic effects we use candidates with decay times larger than 0.6~ps. We also choose an upper  time cut of 7.0~ps, because the acceptance is poorly determined beyond this value.
The acceptance ratio is fitted with a linear function between 0.6 and 7.0 ps. The slope is $a=0.0033\pm0.0024~\rm ps^{-1}$, and the $\chi^2/$number of degrees of freedom (ndf) of the fit is $81/62$.  

\begin{figure}[b]
\begin{center}
    \includegraphics[width=0.75\textwidth]{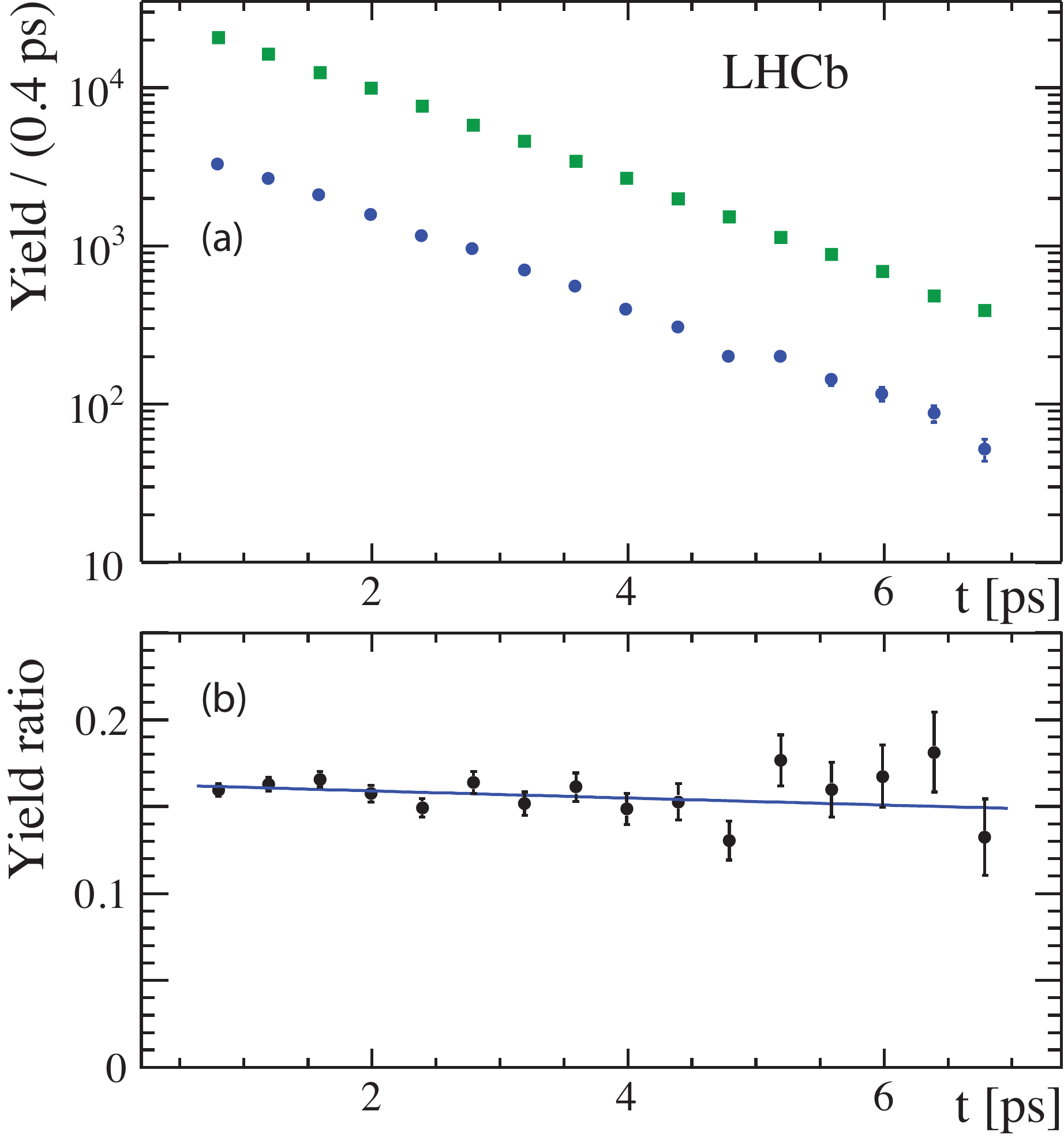}%
\end{center}\label{fig:YieldANDratio}
\vskip -0.8cm
\caption{\small  (a) Decay time distributions for $\Lb \to \jpsi pK^-$ shown as (blue) circles, and  $\Bdb \to \jpsi \Kstarzb(892)$ decays shown as (green) squares. For most entries the error bars are smaller than the points. (b) Yield ratio of $\Lb \to \jpsi pK^-$ to $\Bdb \to \jpsi \Kstarzb(892)$ events fitted as a function of decay time.}
\end{figure}

We determine the event yields in both decay modes by fitting the invariant mass distributions  in 16 bins of decay time, each bin 0.4 ps wide, using the same signal and background shapes obtained in the aforementioned mass fits. Since the bin size is approximately ten times the resolution, there is no effect due to the small difference of time resolution ($<$7\%) between the two modes.
The resulting distributions are shown in Fig.~\ref{fig:YieldANDratio}(a). Here the  fitted signal yields in both modes are placed at the average of the decay time within a bin determined by the \Bzb data in order to correct for
the exponential decrease of the decay time distributions across the bin. 
The decay time ratio distribution fitted with the function given in Eq.~(\ref{eq:raterat2}) is shown in Fig.~\ref{fig:YieldANDratio}(b).
 The $\chi^2/\rm ndf$ of the fit is $18/14$, with a p-value of 21\%. The fitted value of the reciprocal lifetime difference is
\begin{equation*}
\Delta_{\Lz B} = 16.4 \pm 8.2 \pm 4.4~ \rm ns^{-1}.
\end{equation*}
Whenever two uncertainties are quoted, the first is the statistical and the second systematic; the latter will be discussed below.  Numerically,  the ratio of lifetimes is 
\begin{equation*}
\frac{\tau_{\Lb}}{\tau_{\Bzb}}= \frac{1}{1+ \tau_{\Bdb}\Delta_{\Lz B}} 
=0.976\pm0.012\pm0.006, 
\end{equation*}
where we use the world average value $\tau_{\Bzb}=1.519\pm 0.007$~ps \cite{PDG}.
Multiplying the lifetime ratio by this value we determine
\begin{equation*}
\tau_{\Lb}= 1.482 \pm 0.018 \pm 0.012~ \rm ps.
\end{equation*}
Our result is consistent with, but higher and more accurate,  than the current world average of 1.429$\pm$0.024~ps \cite{PDG}.
 
The absolute systematic uncertainties are listed in Table~\ref{tab:sys}.  
\begin{table}[b]
\centering
\caption{\small Absolute systematic uncertainties on $\Delta_{\Lz B}$, the lifetime ratio, and the $\Lb$ lifetime.}
\vspace{0.2cm}
\begin{tabular}{l|cccc}
\hline
Source&$\Delta_{\Lz B}$ (ns$^{-1}$)& $\tau_{\Lb}/\tau_{\Bdb}$ & $\tau_{\Lb}$ (fs) \\
\hline
Decay time fit range &3.2& 0.0045 & 6.9 \\
Acceptance slope &2.3& 0.0033  & 5.0\\
Signal shape & 1.4 & 0.0021 & 3.2 \\
Background model &1.2& 0.0017 & 2.6  \\
$pK$ helicity &0.1& 0.0002 & 0.2\\
Acceptance function  &0.1& 0.0001 & 0.2\\
$\Bdb$ lifetime &-& 0.0001 & 6.8 \\
\hline   
Total &4.4& 0.0062 & 11.7\\
\hline    
\end{tabular}
\label{tab:sys}
\end{table}
There is an uncertainty due to the decay time range used because of the possible change of the acceptance ratio at short decay times. 
This uncertainty is ascertained by changing the fit range to be $1-7$~ps and using the difference with the baseline fit. To determine the acceptance slope uncertainty we vary the value of $a$ by its error determined from the fit to the simulation samples and propagate this change to the results. For the signal shape uncertainty, we repeat the measurement of $\Delta_{\Lz B}$ using a double-Gaussian signal shape in the mass fits. The uncertainty in the background parameterization is assigned by letting the background parameters vary in the fits to the time dependent yields and comparing the difference in final results. Effects of changes in the acceptance for the \Lb mode due to the angular decay distributions are evaluated by weighting the simulation by the observed $pK^-$ helicity angle in addition to  the $pK^-$ invariant mass, and redoing the analysis. The acceptance function uncertainty is evaluated by using a parabola instead of a linear function. The total systematic uncertainty is obtained by adding all of the elements in quadrature.

In conclusion, our value for $\tau_{\Lb}/\tau_{\Bzb} =0.976\pm0.012\pm0.006$ shows that the \Lb and \Bzb lifetimes are indeed equal to within a few percent, as the original advocates of the HQE claimed \cite{Neubert:1996we,Uraltsev:1998bk,Bigi:1995jr,*Bigi:1994wa}, without any need to find additional corrections.  Adding both uncertainties in quadrature, the lifetimes are consistent with being equal at the level of 1.9 standard deviations; thus we do not exclude that the \Lb baryon has a longer lifetime than the \Bzb meson. Using the world average measured value for the \Bzb lifetime we determine
$\tau_{\Lb}= 1.482 \pm 0.018 \pm 0.012$~ ps.

\vspace*{2mm}
We are thankful for many useful and
interesting conversations with Prof. Nikolai Uraltsev who unfortunately
passed away in Feb. 2013, and contributed greatly to theories describing heavy hadron lifetimes.
 We express our gratitude to our colleagues in the CERN
accelerator departments for the excellent performance of the LHC. We
thank the technical and administrative staff at the LHCb
institutes. We acknowledge support from CERN and from the national
agencies: CAPES, CNPq, FAPERJ and FINEP (Brazil); NSFC (China);
CNRS/IN2P3 and Region Auvergne (France); BMBF, DFG, HGF and MPG
(Germany); SFI (Ireland); INFN (Italy); FOM and NWO (The Netherlands);
SCSR (Poland); MEN/IFA (Romania); MinES, Rosatom, RFBR and NRC
``Kurchatov Institute'' (Russia); MinECo, XuntaGal and GENCAT (Spain);
SNSF and SER (Switzerland); NAS Ukraine (Ukraine); STFC (United
Kingdom); NSF (USA). We also acknowledge the support received from the
ERC under FP7. The Tier1 computing centres are supported by IN2P3
(France), KIT and BMBF (Germany), INFN (Italy), NWO and SURF (The
Netherlands), PIC (Spain), GridPP (United Kingdom). We are thankful
for the computing resources put at our disposal by Yandex LLC
(Russia), as well as to the communities behind the multiple open
source software packages that we depend on.

\addcontentsline{toc}{section}{References}
\setboolean{inbibliography}{true}

\begin{mcitethebibliography}{10}
\mciteSetBstSublistMode{n}
\mciteSetBstMaxWidthForm{subitem}{\alph{mcitesubitemcount})}
\mciteSetBstSublistLabelBeginEnd{\mcitemaxwidthsubitemform\space}
{\relax}{\relax}

\bibitem{Kowa-Mannel}
R.~Kowalewski and T.~Mannel, \ifthenelse{\boolean{articletitles}}{{\it
  {Determination of $|V_{ub}|$ and $|V_{cb}|$ in Review of particle physics}},
  }{}\href{http://dx.doi.org/10.1103/PhysRevD.86.010001}{Phys.\ Rev.\  {\bf
  D86} (2012) 010001}\relax
\mciteBstWouldAddEndPuncttrue
\mciteSetBstMidEndSepPunct{\mcitedefaultmidpunct}
{\mcitedefaultendpunct}{\mcitedefaultseppunct}\relax
\EndOfBibitem
\bibitem{Bernlochner:2010zz}
F.~U. Bernlochner {\em et~al.}, \ifthenelse{\boolean{articletitles}}{{\it
  {Towards a global fit to extract the $B\to X_s\gamma$ decay rate and
  $|V_{ub}|$}}, }{}PoS {\bf ICHEP2010} (2010) 229,
  \href{http://arxiv.org/abs/1011.5838}{{\tt arXiv:1011.5838}}\relax
\mciteBstWouldAddEndPuncttrue
\mciteSetBstMidEndSepPunct{\mcitedefaultmidpunct}
{\mcitedefaultendpunct}{\mcitedefaultseppunct}\relax
\EndOfBibitem
\bibitem{Laiho:2012ss}
J.~Laiho, E.~Lunghi, and R.~Van~de Water,
  \ifthenelse{\boolean{articletitles}}{{\it {Flavor physics in the LHC era: the
  role of the lattice}}, }{}PoS {\bf LATTICE2011} (2011) 018,
  \href{http://arxiv.org/abs/1204.0791}{{\tt arXiv:1204.0791}}\relax
\mciteBstWouldAddEndPuncttrue
\mciteSetBstMidEndSepPunct{\mcitedefaultmidpunct}
{\mcitedefaultendpunct}{\mcitedefaultseppunct}\relax
\EndOfBibitem
\bibitem{Stone:2012yr}
S.~Stone, \ifthenelse{\boolean{articletitles}}{{\it {New physics from
  flavour}}, }{}\href{http://arxiv.org/abs/1212.6374}{{\tt
  arXiv:1212.6374}}\relax
\mciteBstWouldAddEndPuncttrue
\mciteSetBstMidEndSepPunct{\mcitedefaultmidpunct}
{\mcitedefaultendpunct}{\mcitedefaultseppunct}\relax
\EndOfBibitem
\bibitem{Bigi:1995jr}
I.~I. Bigi, \ifthenelse{\boolean{articletitles}}{{\it {The QCD perspective on
  lifetimes of heavy flavor hadrons}},
  }{}\href{http://arxiv.org/abs/hep-ph/9508408}{{\tt
  arXiv:hep-ph/9508408}}\relax
\mciteBstWouldAddEndPuncttrue
\mciteSetBstMidEndSepPunct{\mcitedefaultmidpunct}
{\mcitedefaultendpunct}{\mcitedefaultseppunct}\relax
\EndOfBibitem
\bibitem{Bigi:1994wa}
I.~I. Bigi {\em et~al.}, \ifthenelse{\boolean{articletitles}}{{\it {Nonleptonic
  decays of beauty hadrons: from phenomenology to theory}},
  }{}\href{http://arxiv.org/abs/hep-ph/9401298}{{\tt arXiv:hep-ph/9401298}}, in
  {``$B$ Decays,"} edited by S. Stone (World Scientific, Singapore, 1994)\relax
\mciteBstWouldAddEndPuncttrue
\mciteSetBstMidEndSepPunct{\mcitedefaultmidpunct}
{\mcitedefaultendpunct}{\mcitedefaultseppunct}\relax
\EndOfBibitem
\bibitem{Uraltsev:1998bk}
N.~Uraltsev, \ifthenelse{\boolean{articletitles}}{{\it {Heavy quark expansion
  in beauty and its decays}},
  }{}\href{http://arxiv.org/abs/hep-ph/9804275}{{\tt
  arXiv:hep-ph/9804275}}\relax
\mciteBstWouldAddEndPuncttrue
\mciteSetBstMidEndSepPunct{\mcitedefaultmidpunct}
{\mcitedefaultendpunct}{\mcitedefaultseppunct}\relax
\EndOfBibitem
\bibitem{Neubert:1997gu}
M.~Neubert, \ifthenelse{\boolean{articletitles}}{{\it {B decays and the heavy
  quark expansion}}, }{}Adv.\ Ser.\ Direct.\ High Energy Phys.\  {\bf 15}
  (1998) 239, \href{http://arxiv.org/abs/hep-ph/9702375}{{\tt
  arXiv:hep-ph/9702375}}\relax
\mciteBstWouldAddEndPuncttrue
\mciteSetBstMidEndSepPunct{\mcitedefaultmidpunct}
{\mcitedefaultendpunct}{\mcitedefaultseppunct}\relax
\EndOfBibitem
\bibitem{Wilson:1972ee}
K.~Wilson and W.~Zimmermann, \ifthenelse{\boolean{articletitles}}{{\it
  {Operator product expansions and composite field operators in the general
  framework of quantum field theory}},
  }{}\href{http://dx.doi.org/10.1007/BF01878448}{Commun.\ Math.\ Phys.\  {\bf
  24} (1972) 87}\relax
\mciteBstWouldAddEndPuncttrue
\mciteSetBstMidEndSepPunct{\mcitedefaultmidpunct}
{\mcitedefaultendpunct}{\mcitedefaultseppunct}\relax
\EndOfBibitem
\bibitem{Buchalla:1995vs}
G.~Buchalla, A.~J. Buras, and M.~E. Lautenbacher,
  \ifthenelse{\boolean{articletitles}}{{\it {Weak decays beyond leading
  logarithms}}, }{}\href{http://dx.doi.org/10.1103/RevModPhys.68.1125}{Rev.\
  Mod.\ Phys.\  {\bf 68} (1996) 1125},
  \href{http://arxiv.org/abs/hep-ph/9512380}{{\tt arXiv:hep-ph/9512380}}\relax
\mciteBstWouldAddEndPuncttrue
\mciteSetBstMidEndSepPunct{\mcitedefaultmidpunct}
{\mcitedefaultendpunct}{\mcitedefaultseppunct}\relax
\EndOfBibitem
\bibitem{Falk:2000tx}
A.~F. Falk, \ifthenelse{\boolean{articletitles}}{{\it {The CKM matrix and the
  heavy quark expansion}}, }{}\href{http://arxiv.org/abs/hep-ph/0007339}{{\tt
  arXiv:hep-ph/0007339}}\relax
\mciteBstWouldAddEndPuncttrue
\mciteSetBstMidEndSepPunct{\mcitedefaultmidpunct}
{\mcitedefaultendpunct}{\mcitedefaultseppunct}\relax
\EndOfBibitem
\bibitem{Buras:2011we}
A.~J. Buras, \ifthenelse{\boolean{articletitles}}{{\it {Climbing NLO and NNLO
  summits of weak decays}}, }{}\href{http://arxiv.org/abs/1102.5650}{{\tt
  arXiv:1102.5650}}\relax
\mciteBstWouldAddEndPuncttrue
\mciteSetBstMidEndSepPunct{\mcitedefaultmidpunct}
{\mcitedefaultendpunct}{\mcitedefaultseppunct}\relax
\EndOfBibitem
\bibitem{Cheng:1997xba}
H.-Y. Cheng, \ifthenelse{\boolean{articletitles}}{{\it {Phenomenological
  analysis of heavy hadron lifetimes}},
  }{}\href{http://dx.doi.org/10.1103/PhysRevD.56.2783}{Phys.\ Rev.\  {\bf D56}
  (1997) 2783}, \href{http://arxiv.org/abs/hep-ph/9704260}{{\tt
  arXiv:hep-ph/9704260}}\relax
\mciteBstWouldAddEndPuncttrue
\mciteSetBstMidEndSepPunct{\mcitedefaultmidpunct}
{\mcitedefaultendpunct}{\mcitedefaultseppunct}\relax
\EndOfBibitem
\bibitem{Neubert:1996we}
M.~Neubert and C.~T. Sachrajda, \ifthenelse{\boolean{articletitles}}{{\it
  {Spectator effects in inclusive decays of beauty hadrons}},
  }{}\href{http://dx.doi.org/10.1016/S0550-3213(96)00559-7}{Nucl.\ Phys.\  {\bf
  B483} (1997) 339}, \href{http://arxiv.org/abs/hep-ph/9603202}{{\tt
  arXiv:hep-ph/9603202}}\relax
\mciteBstWouldAddEndPuncttrue
\mciteSetBstMidEndSepPunct{\mcitedefaultmidpunct}
{\mcitedefaultendpunct}{\mcitedefaultseppunct}\relax
\EndOfBibitem
\bibitem{Uraltsev:1996ta}
N.~Uraltsev, \ifthenelse{\boolean{articletitles}}{{\it {On the problem of
  boosting nonleptonic b baryon decays}},
  }{}\href{http://dx.doi.org/10.1016/0370-2693(96)00305-X}{Phys.\ Lett.\  {\bf
  B376} (1996) 303}, \href{http://arxiv.org/abs/hep-ph/9602324}{{\tt
  arXiv:hep-ph/9602324}}\relax
\mciteBstWouldAddEndPuncttrue
\mciteSetBstMidEndSepPunct{\mcitedefaultmidpunct}
{\mcitedefaultendpunct}{\mcitedefaultseppunct}\relax
\EndOfBibitem
\bibitem{DiPierro:1999tb}
UKQCD collaboration, M.~Di~Pierro, C.~T. Sachrajda, and C.~Michael,
  \ifthenelse{\boolean{articletitles}}{{\it {An Exploratory lattice study of
  spectator effects in inclusive decays of the Lambda(b) baryon}},
  }{}\href{http://dx.doi.org/10.1016/S0370-2693(99)01166-1}{Phys.\ Lett.\  {\bf
  B468} (1999) 143}, \href{http://arxiv.org/abs/hep-lat/9906031}{{\tt
  arXiv:hep-lat/9906031}}\relax
\mciteBstWouldAddEndPuncttrue
\mciteSetBstMidEndSepPunct{\mcitedefaultmidpunct}
{\mcitedefaultendpunct}{\mcitedefaultseppunct}\relax
\EndOfBibitem
\bibitem{Rosner:1996fy}
J.~L. Rosner, \ifthenelse{\boolean{articletitles}}{{\it {Enhancement of the \Lb
  decay rate}}, }{}\href{http://dx.doi.org/10.1016/0370-2693(96)00352-8}{Phys.\
  Lett.\  {\bf B379} (1996) 267},
  \href{http://arxiv.org/abs/hep-ph/9602265}{{\tt arXiv:hep-ph/9602265}}\relax
\mciteBstWouldAddEndPuncttrue
\mciteSetBstMidEndSepPunct{\mcitedefaultmidpunct}
{\mcitedefaultendpunct}{\mcitedefaultseppunct}\relax
\EndOfBibitem
\bibitem{Battaglia:2003in}
M.~Battaglia {\em et~al.}, \ifthenelse{\boolean{articletitles}}{{\it {The CKM
  matrix and the unitarity triangle}},
  }{}\href{http://arxiv.org/abs/hep-ph/0304132}{{\tt
  arXiv:hep-ph/0304132}}\relax
\mciteBstWouldAddEndPuncttrue
\mciteSetBstMidEndSepPunct{\mcitedefaultmidpunct}
{\mcitedefaultendpunct}{\mcitedefaultseppunct}\relax
\EndOfBibitem
\bibitem{Tarantino:2003qw}
C.~Tarantino, \ifthenelse{\boolean{articletitles}}{{\it {Beauty hadron
  lifetimes and B meson \CP violation parameters from lattice QCD}},
  }{}\href{http://dx.doi.org/10.1140/epjcd/s2003-03-1006-y}{Eur.\ Phys.\ J.\
  {\bf C33} (2004) S895}, \href{http://arxiv.org/abs/hep-ph/0310241}{{\tt
  arXiv:hep-ph/0310241}}\relax
\mciteBstWouldAddEndPuncttrue
\mciteSetBstMidEndSepPunct{\mcitedefaultmidpunct}
{\mcitedefaultendpunct}{\mcitedefaultseppunct}\relax
\EndOfBibitem
\bibitem{Franco:2002fc}
E.~Franco, V.~Lubicz, F.~Mescia, and C.~Tarantino,
  \ifthenelse{\boolean{articletitles}}{{\it {Lifetime ratios of beauty hadrons
  at the next-to-leading order in QCD}},
  }{}\href{http://dx.doi.org/10.1016/S0550-3213(02)00262-6}{Nucl.\ Phys.\  {\bf
  B633} (2002) 212}, \href{http://arxiv.org/abs/hep-ph/0203089}{{\tt
  arXiv:hep-ph/0203089}}\relax
\mciteBstWouldAddEndPuncttrue
\mciteSetBstMidEndSepPunct{\mcitedefaultmidpunct}
{\mcitedefaultendpunct}{\mcitedefaultseppunct}\relax
\EndOfBibitem
\bibitem{Ito:1997qq}
T.~Ito, M.~Matsuda, and Y.~Matsui, \ifthenelse{\boolean{articletitles}}{{\it
  {New possibility of solving the problem of the lifetime ratio
  $\tau(\Lb)/\tau(B_d)$}},
  }{}\href{http://dx.doi.org/10.1143/PTP.99.271}{Prog.\ Theor.\ Phys.\  {\bf
  99} (1998) 271}, \href{http://arxiv.org/abs/hep-ph/9705402}{{\tt
  arXiv:hep-ph/9705402}}\relax
\mciteBstWouldAddEndPuncttrue
\mciteSetBstMidEndSepPunct{\mcitedefaultmidpunct}
{\mcitedefaultendpunct}{\mcitedefaultseppunct}\relax
\EndOfBibitem
\bibitem{Gabbiani:2003pq}
F.~Gabbiani, A.~I. Onishchenko, and A.~A. Petrov,
  \ifthenelse{\boolean{articletitles}}{{\it {\Lb lifetime puzzle in heavy quark
  expansion}}, }{}\href{http://dx.doi.org/10.1103/PhysRevD.68.114006}{Phys.\
  Rev.\  {\bf D68} (2003) 114006},
  \href{http://arxiv.org/abs/hep-ph/0303235}{{\tt arXiv:hep-ph/0303235}}\relax
\mciteBstWouldAddEndPuncttrue
\mciteSetBstMidEndSepPunct{\mcitedefaultmidpunct}
{\mcitedefaultendpunct}{\mcitedefaultseppunct}\relax
\EndOfBibitem
\bibitem{Gabbiani:2004tp}
F.~Gabbiani, A.~I. Onishchenko, and A.~A. Petrov,
  \ifthenelse{\boolean{articletitles}}{{\it {Spectator effects and lifetimes of
  heavy hadrons}},
  }{}\href{http://dx.doi.org/10.1103/PhysRevD.70.094031}{Phys.\ Rev.\  {\bf
  D70} (2004) 094031}, \href{http://arxiv.org/abs/hep-ph/0407004}{{\tt
  arXiv:hep-ph/0407004}}\relax
\mciteBstWouldAddEndPuncttrue
\mciteSetBstMidEndSepPunct{\mcitedefaultmidpunct}
{\mcitedefaultendpunct}{\mcitedefaultseppunct}\relax
\EndOfBibitem
\bibitem{Uraltsev:2000qw}
N.~Uraltsev, \ifthenelse{\boolean{articletitles}}{{\it {Topics in the heavy
  quark expansion}}, }{}\href{http://arxiv.org/abs/hep-ph/0010328}{{\tt
  arXiv:hep-ph/0010328}}\relax
\mciteBstWouldAddEndPuncttrue
\mciteSetBstMidEndSepPunct{\mcitedefaultmidpunct}
{\mcitedefaultendpunct}{\mcitedefaultseppunct}\relax
\EndOfBibitem
\bibitem{Aad:2012bpa}
ATLAS collaboration, G.~Aad {\em et~al.},
  \ifthenelse{\boolean{articletitles}}{{\it {Measurement of the $\Lb$ lifetime
  and mass in the ATLAS experiment}},
  }{}\href{http://dx.doi.org/10.1103/PhysRevD.87.032002}{Phys.\ Rev.\  {\bf
  D87} (2013) 032002}, \href{http://arxiv.org/abs/1207.2284}{{\tt
  arXiv:1207.2284}}\relax
\mciteBstWouldAddEndPuncttrue
\mciteSetBstMidEndSepPunct{\mcitedefaultmidpunct}
{\mcitedefaultendpunct}{\mcitedefaultseppunct}\relax
\EndOfBibitem
\bibitem{Chatrchyan:2013sxa}
CMS collaboration, S.~Chatrchyan {\em et~al.},
  \ifthenelse{\boolean{articletitles}}{{\it {Measurement of the $\Lb$ lifetime
  in $pp$ collisions at $\sqrt{s}$ = 7 TeV}},
  }{}\href{http://arxiv.org/abs/1304.7495}{{\tt arXiv:1304.7495}}\relax
\mciteBstWouldAddEndPuncttrue
\mciteSetBstMidEndSepPunct{\mcitedefaultmidpunct}
{\mcitedefaultendpunct}{\mcitedefaultseppunct}\relax
\EndOfBibitem
\bibitem{Aaltonen:2009zn}
CDF collaboration, T.~Aaltonen {\em et~al.},
  \ifthenelse{\boolean{articletitles}}{{\it {Measurement of the \Lb lifetime in
  $\Lb\to\Lc\pi^-$ decays in $p\overline{p}$ collisions at $\sqrt{s}$ = 1.96
  TeV}}, }{}\href{http://dx.doi.org/10.1103/PhysRevLett.104.102002}{Phys.\
  Rev.\ Lett.\  {\bf 104} (2010) 102002},
  \href{http://arxiv.org/abs/0912.3566}{{\tt arXiv:0912.3566}}\relax
\mciteBstWouldAddEndPuncttrue
\mciteSetBstMidEndSepPunct{\mcitedefaultmidpunct}
{\mcitedefaultendpunct}{\mcitedefaultseppunct}\relax
\EndOfBibitem
\bibitem{Aaltonen:2010pj}
CDF collaboration, T.~Aaltonen {\em et~al.},
  \ifthenelse{\boolean{articletitles}}{{\it {Measurement of $b$ hadron
  lifetimes in exclusive decays containing a $J/\psi$ in $p\overline{p}$
  collisions at $\sqrt{s}=1.96$~TeV}},
  }{}\href{http://dx.doi.org/10.1103/PhysRevLett.106.121804}{Phys.\ Rev.\
  Lett.\  {\bf 106} (2011) 121804}, \href{http://arxiv.org/abs/1012.3138}{{\tt
  arXiv:1012.3138}}\relax
\mciteBstWouldAddEndPuncttrue
\mciteSetBstMidEndSepPunct{\mcitedefaultmidpunct}
{\mcitedefaultendpunct}{\mcitedefaultseppunct}\relax
\EndOfBibitem
\bibitem{Abazov:2012iy}
D0 collaboration, V.~M. Abazov {\em et~al.},
  \ifthenelse{\boolean{articletitles}}{{\it {Measurement of the $\Lb$ lifetime
  in the exclusive decay $\Lb \to J/\psi \L$ in $p\bar{p}$ collisions at
  $\sqrt{s}=1.96$ TeV}},
  }{}\href{http://dx.doi.org/10.1103/PhysRevD.85.112003}{Phys.\ Rev.\  {\bf
  D85} (2012) 112003}, \href{http://arxiv.org/abs/1204.2340}{{\tt
  arXiv:1204.2340}}\relax
\mciteBstWouldAddEndPuncttrue
\mciteSetBstMidEndSepPunct{\mcitedefaultmidpunct}
{\mcitedefaultendpunct}{\mcitedefaultseppunct}\relax
\EndOfBibitem
\bibitem{LHCb-det}
LHCb collaboration, A.~Alves~Jr. {\em et~al.},
  \ifthenelse{\boolean{articletitles}}{{\it {The LHCb detector at the LHC}},
  }{}\href{http://dx.doi.org/10.1088/1748-0221/3/08/S08005}{JINST {\bf 3}
  (2008) S08005}\relax
\mciteBstWouldAddEndPuncttrue
\mciteSetBstMidEndSepPunct{\mcitedefaultmidpunct}
{\mcitedefaultendpunct}{\mcitedefaultseppunct}\relax
\EndOfBibitem
\bibitem{Aaij:2012me}
R.~Aaij {\em et~al.}, \ifthenelse{\boolean{articletitles}}{{\it {The LHCb
  trigger and its performance in 2011}},
  }{}\href{http://dx.doi.org/10.1088/1748-0221/8/04/P04022}{JINST {\bf 8}
  (2013) P04022}, \href{http://arxiv.org/abs/1211.3055}{{\tt
  arXiv:1211.3055}}\relax
\mciteBstWouldAddEndPuncttrue
\mciteSetBstMidEndSepPunct{\mcitedefaultmidpunct}
{\mcitedefaultendpunct}{\mcitedefaultseppunct}\relax
\EndOfBibitem
\bibitem{Sjostrand:2006za}
T.~Sj{\"o}strand, S.~Mrenna, and P.~Z. Skands,
  \ifthenelse{\boolean{articletitles}}{{\it {PYTHIA 6.4 physics and manual}},
  }{}\href{http://dx.doi.org/10.1088/1126-6708/2006/05/026}{JHEP {\bf 05}
  (2006) 026}, \href{http://arxiv.org/abs/hep-ph/0603175}{{\tt
  arXiv:hep-ph/0603175}}\relax
\mciteBstWouldAddEndPuncttrue
\mciteSetBstMidEndSepPunct{\mcitedefaultmidpunct}
{\mcitedefaultendpunct}{\mcitedefaultseppunct}\relax
\EndOfBibitem
\bibitem{LHCb-PROC-2011-005}
I.~Belyaev {\em et~al.}, \ifthenelse{\boolean{articletitles}}{{\it {Handling of
  the generation of primary events in \gauss, the \lhcb simulation framework}},
  }{}\href{http://dx.doi.org/10.1109/NSSMIC.2010.5873949}{Nuclear Science
  Symposium Conference Record (NSS/MIC) {\bf IEEE} (2010) 1155}\relax
\mciteBstWouldAddEndPuncttrue
\mciteSetBstMidEndSepPunct{\mcitedefaultmidpunct}
{\mcitedefaultendpunct}{\mcitedefaultseppunct}\relax
\EndOfBibitem
\bibitem{Agostinelli:2002hh}
GEANT4 collaboration, S.~Agostinelli {\em et~al.},
  \ifthenelse{\boolean{articletitles}}{{\it {GEANT4 - a simulation toolkit}},
  }{}\href{http://dx.doi.org/10.1016/S0168-9002(03)01368-8}{Nucl.\ Instrum.\
  Meth.\  {\bf A506} (2003) 250}\relax
\mciteBstWouldAddEndPuncttrue
\mciteSetBstMidEndSepPunct{\mcitedefaultmidpunct}
{\mcitedefaultendpunct}{\mcitedefaultseppunct}\relax
\EndOfBibitem
\bibitem{Allison:2006ve}
GEANT4 collaboration, J.~Allison {\em et~al.},
  \ifthenelse{\boolean{articletitles}}{{\it {Geant4 developments and
  applications}}, }{}\href{http://dx.doi.org/10.1109/TNS.2006.869826}{IEEE
  Trans.\ Nucl.\ Sci.\  {\bf 53} (2006) 270}\relax
\mciteBstWouldAddEndPuncttrue
\mciteSetBstMidEndSepPunct{\mcitedefaultmidpunct}
{\mcitedefaultendpunct}{\mcitedefaultseppunct}\relax
\EndOfBibitem
\bibitem{LHCb-PROC-2011-006}
M.~Clemencic {\em et~al.}, \ifthenelse{\boolean{articletitles}}{{\it {The \lhcb
  simulation application, \gauss: design, evolution and experience}},
  }{}\href{http://dx.doi.org/10.1088/1742-6596/331/3/032023}{{J.\ Phys.\ \!\!:
  Conf.\ Ser.\ } {\bf 331} (2011) 032023}\relax
\mciteBstWouldAddEndPuncttrue
\mciteSetBstMidEndSepPunct{\mcitedefaultmidpunct}
{\mcitedefaultendpunct}{\mcitedefaultseppunct}\relax
\EndOfBibitem
\bibitem{Lange:2001uf}
D.~Lange, \ifthenelse{\boolean{articletitles}}{{\it {The EvtGen particle decay
  simulation package}},
  }{}\href{http://dx.doi.org/10.1016/S0168-9002(01)00089-4}{Nucl.\ Instrum.\
  Meth.\  {\bf A462} (2001) 152}\relax
\mciteBstWouldAddEndPuncttrue
\mciteSetBstMidEndSepPunct{\mcitedefaultmidpunct}
{\mcitedefaultendpunct}{\mcitedefaultseppunct}\relax
\EndOfBibitem
\bibitem{Breiman}
L.~Breiman, J.~H. Friedman, R.~A. Olshen, and C.~J. Stone, {\em Classification
  and regression trees}, Wadsworth international group, Belmont, California,
  USA, 1984\relax
\mciteBstWouldAddEndPuncttrue
\mciteSetBstMidEndSepPunct{\mcitedefaultmidpunct}
{\mcitedefaultendpunct}{\mcitedefaultseppunct}\relax
\EndOfBibitem
\bibitem{Adinolfi:2012an}
M.~Adinolfi {\em et~al.}, \ifthenelse{\boolean{articletitles}}{{\it
  {Performance of the LHCb RICH detector at the LHC}},
  }{}\href{http://dx.doi.org/10.1140/epjc/s10052-013-2431-9}{Eur.\ Phys.\ J.\
  {\bf C73} (2013) 2431}, \href{http://arxiv.org/abs/1211.6759}{{\tt
  arXiv:1211.6759}}\relax
\mciteBstWouldAddEndPuncttrue
\mciteSetBstMidEndSepPunct{\mcitedefaultmidpunct}
{\mcitedefaultendpunct}{\mcitedefaultseppunct}\relax
\EndOfBibitem
\bibitem{Aaij:2013orb}
LHCb collaboration, R.~Aaij {\em et~al.},
  \ifthenelse{\boolean{articletitles}}{{\it {Amplitude analysis and branching
  fraction measurement of $\Bsb\to\jpsi\KpKm$}},
  }{}\href{http://dx.doi.org/10.1103/PhysRevD.87.072004}{Phys.\ Rev.\  {\bf
  D87} (2013) 072004}, \href{http://arxiv.org/abs/1302.1213}{{\tt
  arXiv:1302.1213}}\relax
\mciteBstWouldAddEndPuncttrue
\mciteSetBstMidEndSepPunct{\mcitedefaultmidpunct}
{\mcitedefaultendpunct}{\mcitedefaultseppunct}\relax
\EndOfBibitem
\bibitem{PDG}
Particle Data Group, J.~Beringer {\em et~al.},
  \ifthenelse{\boolean{articletitles}}{{\it {Review of particle physics and
  2013 partial update for the 2014 edition. }},
  }{}\href{http://dx.doi.org/10.1103/PhysRevD.86.010001}{Phys.\ Rev.\  {\bf
  D86} (2012) 010001}\relax
\mciteBstWouldAddEndPuncttrue
\mciteSetBstMidEndSepPunct{\mcitedefaultmidpunct}
{\mcitedefaultendpunct}{\mcitedefaultseppunct}\relax
\EndOfBibitem
\end{mcitethebibliography}
\ifx\mcitethebibliography\mciteundefinedmacro
\PackageError{LHCb.bst}{mciteplus.sty has not been loaded}
{This bibstyle requires the use of the mciteplus package.}\fi
\providecommand{\href}[2]{#2}

\end{document}